\documentclass{epl}

\usepackage{graphicx}

\title{From sand to networks: a study of multi-disciplinarity}

\author{R. Lambiotte\inst{1} and M. Ausloos\inst{1}}

\institute{ \inst{1}SUPRATECS, Universit\'e de Li\`ege, B5 Sart-Tilman, B-4000 Li\`ege, Belgium}


\begin{document}

\maketitle

\begin{abstract}
In this paper, we study empirically co-authorship networks of neighbouring scientific disciplines, and describe the system by two coupled networks. By considering a large time window, we focus on the properties of the interface between the disciplines. We also focus on the time evolution of the co-authorship network, and highlight a rich phenomenology including first order transition and cluster bouncing and merging. Finally, we present a simple Ising-like model (CDIM) that reproduces qualitatively the structuring of the system in homogeneous phases.
\end{abstract}

\section{Introduction}

Since the pioneering works of Barabasi and Albert \cite{albert, barabasi},  {\em "complex networks"} have become a more and more active field, attracting physicists from the whole sub-fields of statistical physics, ranging from theoretical non-equilibrium statistical physics to experimental granular compaction. 
These complex structures are usually composed by large number of internal components (the nodes), and describe a wide variety 
of systems of high technological and intellectual importance,
examples including the Internet \cite{internet},  business relations between companies \cite{business}, ecological networks \cite{ecological} and airplane route networks \cite{airplane}. As a paradigm for large-scale social networks, people usually consider co-authorship networks  \cite{newman}, namely networks where nodes represent
scientists, and where a link is drawn between them if they co-authored a common paper. Their study has been very active recently, due to their complex social structure \cite{newman3},  to the ubiquity of their bipartite structure in complex systems \cite{bara} \cite{ramasco}, and to the large databases available (arXiv and Science Index). 

In this paper, we
analyze data for such collaboration networks and focus on the development of neighbouring scientific disciplines in the course of time, thereby eyeing the spreading of new ideas in the science community. 
Let us stress that the identification of the mechanisms responsible for knowledge diffusion and, possibly, 
scientific avalanches, is primordial in order to understand the scientific response to external political decisions, and to develop efficient policy recommendations. 
In section 2, we concentrate empirically on this issue by studying data extracted from the arXiv database. To do so, we discriminate two sub-communities of physicists, those studying {\em "complex networks"} and those studying {\em "granular media"}. This choice is motivated by the relative closeness of these fields, that allows interactions between sub-communities (inter-disciplinarity collaboration), and the passage of a scientist from one field to the other (scientist mobility). The data analysis highlights that most contacts between the two disciplines are driven by inter-disciplinary collaborations, and reveals complex time-dependent properties.
In section 3, we present a simple model based on the empirical observations. It is important to point that science spreading is usually modeled by master equations with auto-catalytic processes \cite{andrea}, or by epidemic models on static networks \cite{holyst}. In this article, however, we present a novel approach where the structure of the network itself evolves in an epidemic way. 

\section {Empirical data}

\begin{figure}
\hspace{1.8cm}
\includegraphics[width=3.5in]{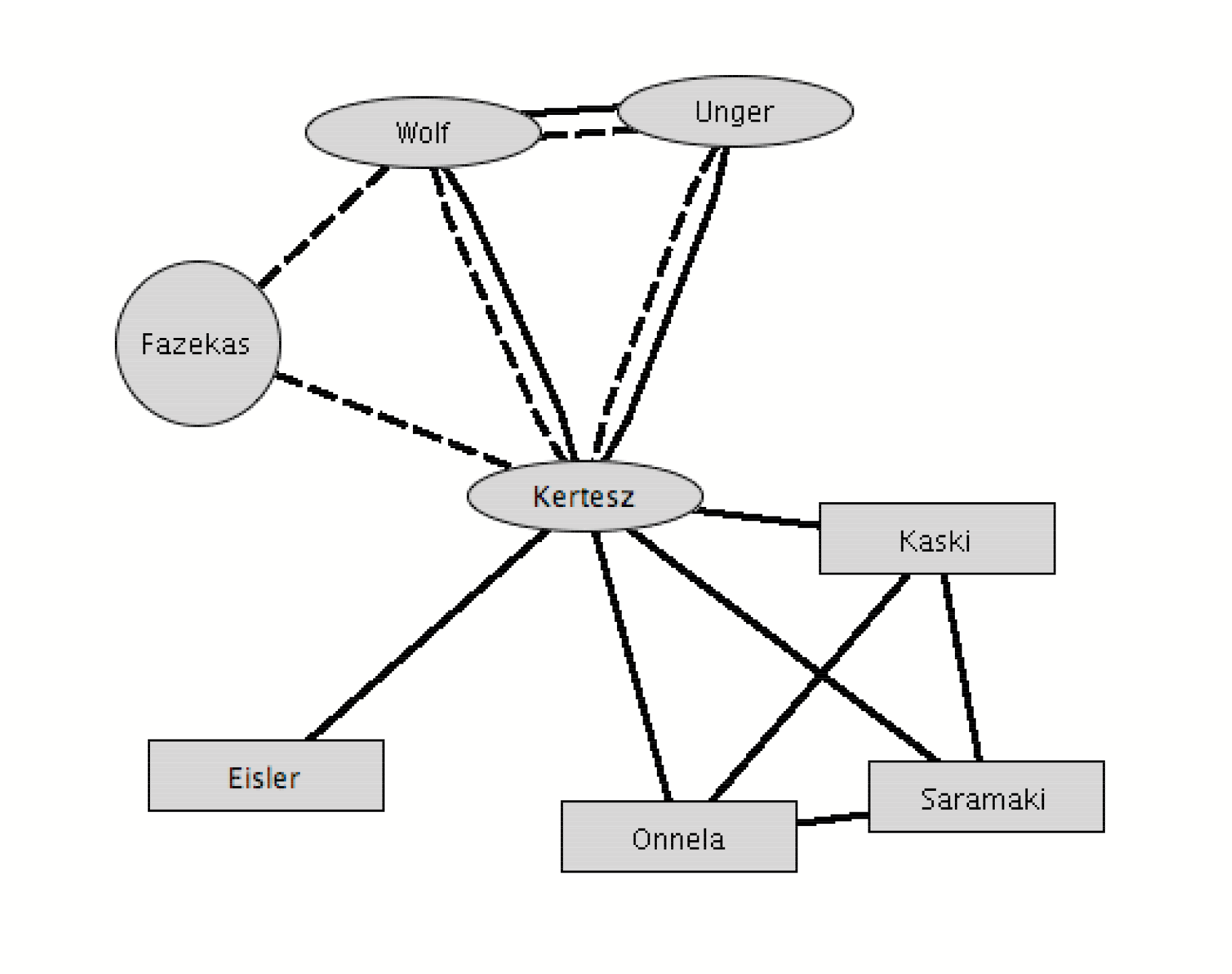}
\caption{\label{little} Small disconnected island from data collected between March 2004 and March 2005. Dashed/Solid lines represent a collaboration in {\em granular media}/{\em networks}. In that system, J. Kertesz plays a central and inter-disciplinary role.}
\end{figure}

\begin{figure}
\hspace{1.8cm}
\includegraphics[width=3.50in]{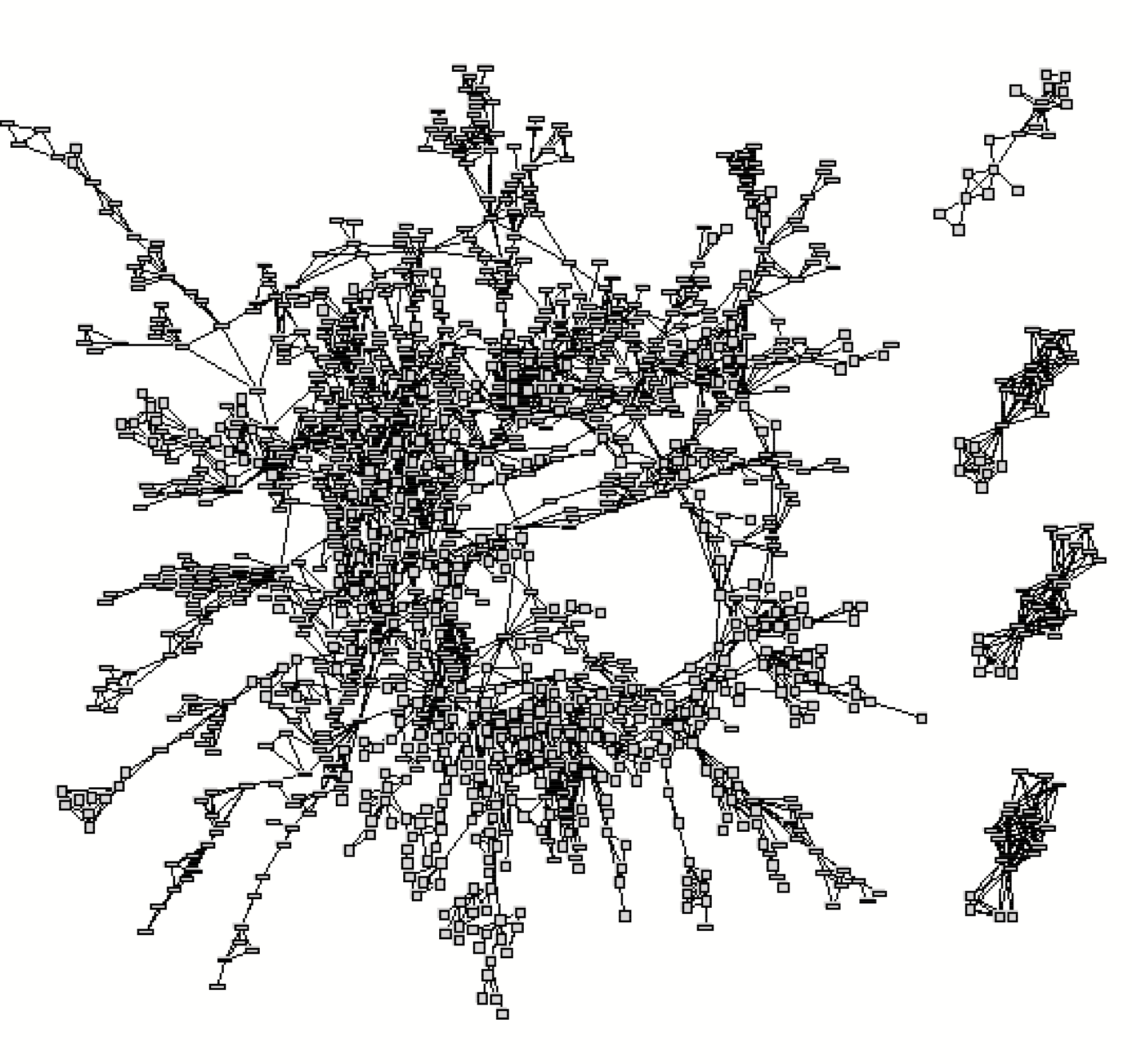}
\caption{\label{10Years} 5 main islands from the data collected between March 1995 and March 2005. The large connected island encompasses  $36\%$ of the total number of scientists.  }
\end{figure}

The data set contains all articles from arXiv in the time interval $[1995:2005]$, that contain the word {\em "network"}  or (exclusive "or") the word {\em "granular"} in their abstract and are classified as {\em "cond-mat"}. In the following, we assume that this simple semantic filter is sufficient to distinguish the collaborations between scientists. Nonetheless, we recognize that the method does not ensure a perfect characterisation of the papers subject, e.g. some {\em "network papers"} may not focus on complex networks, such as {\em The response function of a sphere in a viscoelastic two-fluid medium} by 
Levine and Lubensky \cite{levine}. In order to discriminate the authors and avoid spurious data, we checked the names and the first names of the authors. Moreover, in order to avoid multiple ways for an author to cosign a paper, we also took into account the initial notation of the prenames. For instance, {\em Marcel Ausloos} and {\em M. Ausloos} are the same person, while {\em Marcel Ausloos} and {\em Mike Ausloos} are considered to be different. Let us stress that this method may lead to ambiguities if an initial refers to two different first names. Nonetheless, we have verified that this case occurs only once in the data set (Hawoong, Hyeong-Chai and H. Jeong), so that its effects are negligible. Given this identification method,  we find $3297$ scientists and $2305$ articles. Among these scientists, 105 have written their articles by themselves, i.e. without co-author. As these people are excluded from the co-authorship network, we neglect them in the following. In the $3192$ remaining scientists, 2270 ones have written at least one "network" article, and $1072$ ones have written at least one "granular" article. The $150$ scientists who have written articles in the two fields are obviously multi-disciplinary scientists, and thereby ensure direct communication between the two scientific fields. 

\begin{figure}
\hspace{1.8cm}
\includegraphics[width=3.5in]{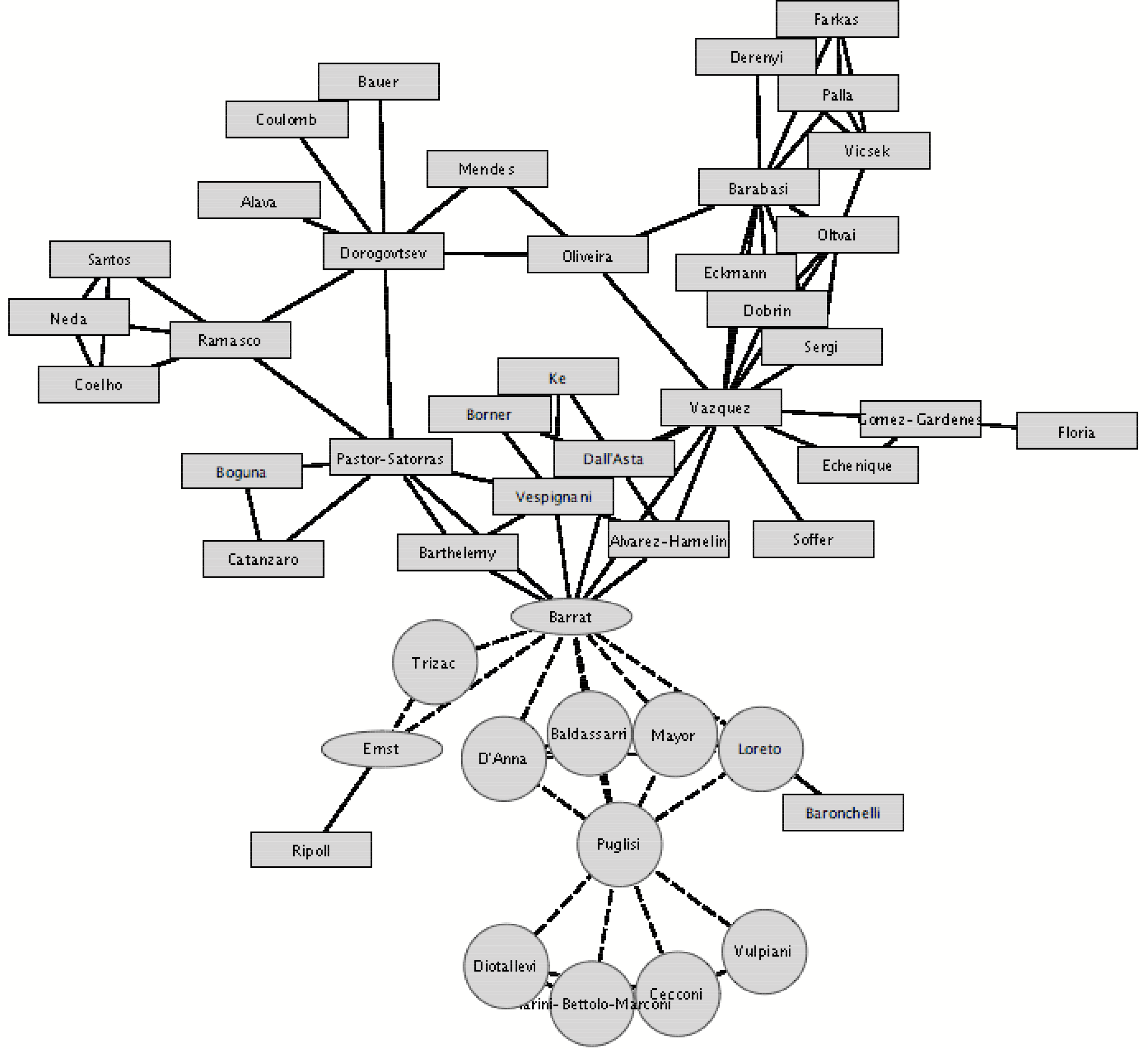}
\caption{\label{mechanism} Zoom on the main island from data collected between March 2004 and March 2005. The system shows two well-separated homogeneous phases, that are connected by one multi-disciplinary scientist (A. Barrat). It is important to note that, if this author is active in both fields, his collaboration networks are distinct in each field.  }
\end{figure}

In order to build the co-authorship network, we apply the usual method \cite{newman2b}, namely we consider a network of scientists placed at nodes, with a link between them if they co-authored a common paper. In order to discriminate the collaboration type between scientists (Fig.\ref{little}), we use different edge shapes, namely solid and dashed lines. Moreover, we also discriminate scientists as {\em "granular"}, {\em "network"} or {\em"multi-disciplinary"} scientists, depending on their collaborations. By convention, if more than $70 \%$ of their links are {\em"granular"}/{\em"network"}, the author is considered as {\em"granular"}/{\em "network"}, and is depicted by a circle/rectangle. Else, the author is multi-disciplinary and is depicted by an ellipse. For a color-based version of the discrimination, see \cite{site}. As a result (Fig.\ref{10Years}), the system is made of two coupled networks, i.e. different networks layed on the same nodes. It  consists in a large 
connected island  of $1180$ scientists and of a multitude of small disconnected clusters, 
--reminiscent of the clique habits of  authors. Let us stress that the main island  exhibits typical features of social networks, e.g.  strongly connected scientists \cite{newman2}  and modular structures \cite{scc}.
In order to study the interface between the two scientific fields, we have focused on the $150$ multi-disciplinary scientists $i$ in the system, each of them being characterized by $N_G^i$ and $N_N^i$ {\em"granular"}/{\em "network"} links. Data analysis shows that their total number of links $N=\frac{1}{2}\sum_{i=1}^{50} N_G^i + N_N^i$ is equal to $733$. In contrast, the total number of collaboration pairs $(i, j)$ that are related by {\bf both} kinds of links is equal to $63$, i.e. there are $126$ such links. 
This shows that most contacts between the two disciplines are driven by a change of the collaboration network (Fig.\ref{mechanism}), and not by a cooperative switch of the collaboration network. In other word, when a scientist works in two fields, he works in each field with different persons. Let us stress, however, that there are notable exceptions, such as the triplet made of {\em (F. Coppex, M. Droz, A. Lipowski)} who work actively together in both fields. It is also important to note the existence of well-defined phases, namely regions of the network homogeneously connected by {\em"granular"}/{\em "network"} links, thereby confirming that authors collaborate primarily with others with whom their research focus is aligned \cite{girvan1}.

\begin{figure}
\hspace{-0.5cm}
\includegraphics[width=2.90in]{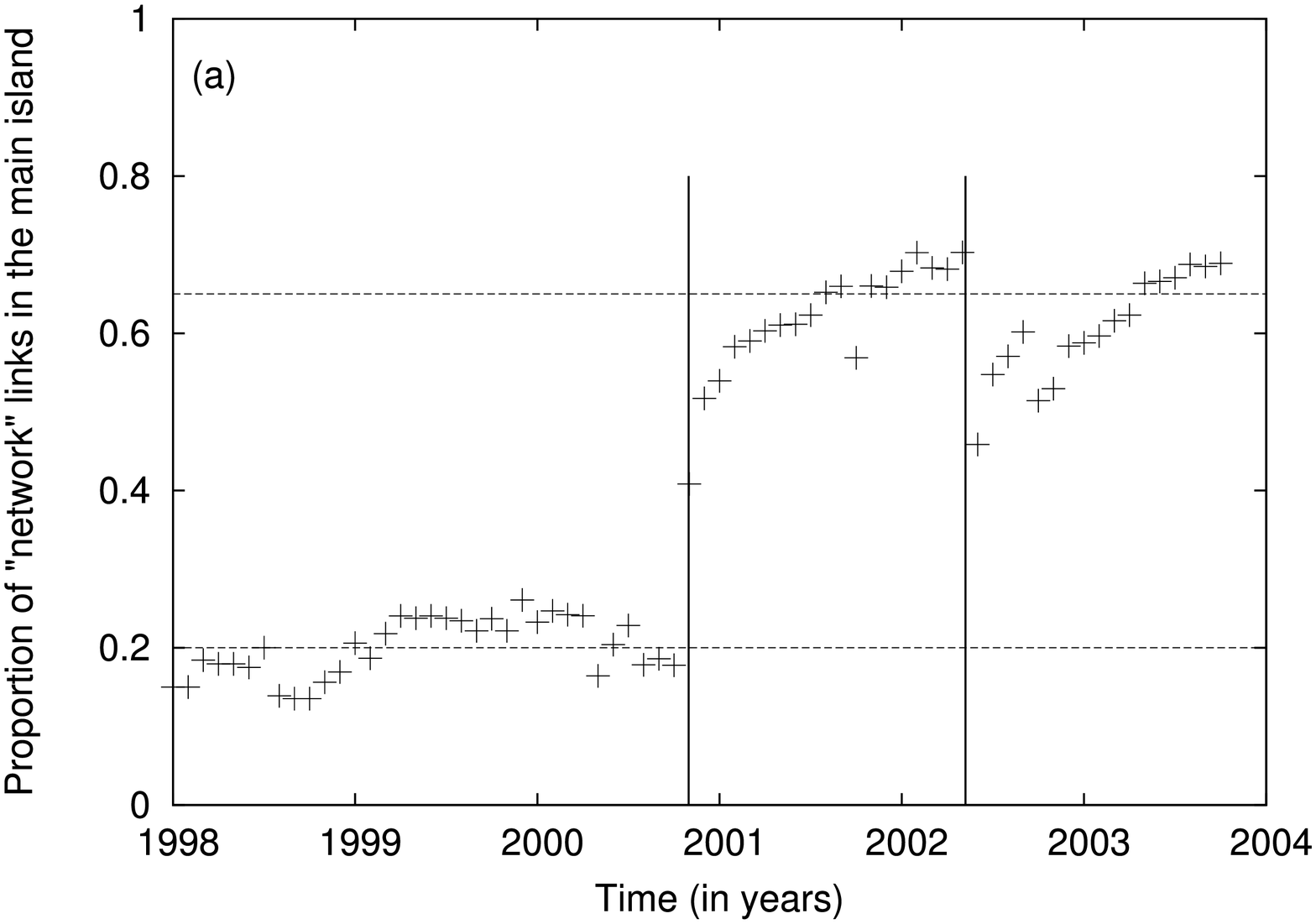}
\includegraphics[width=2.90in]{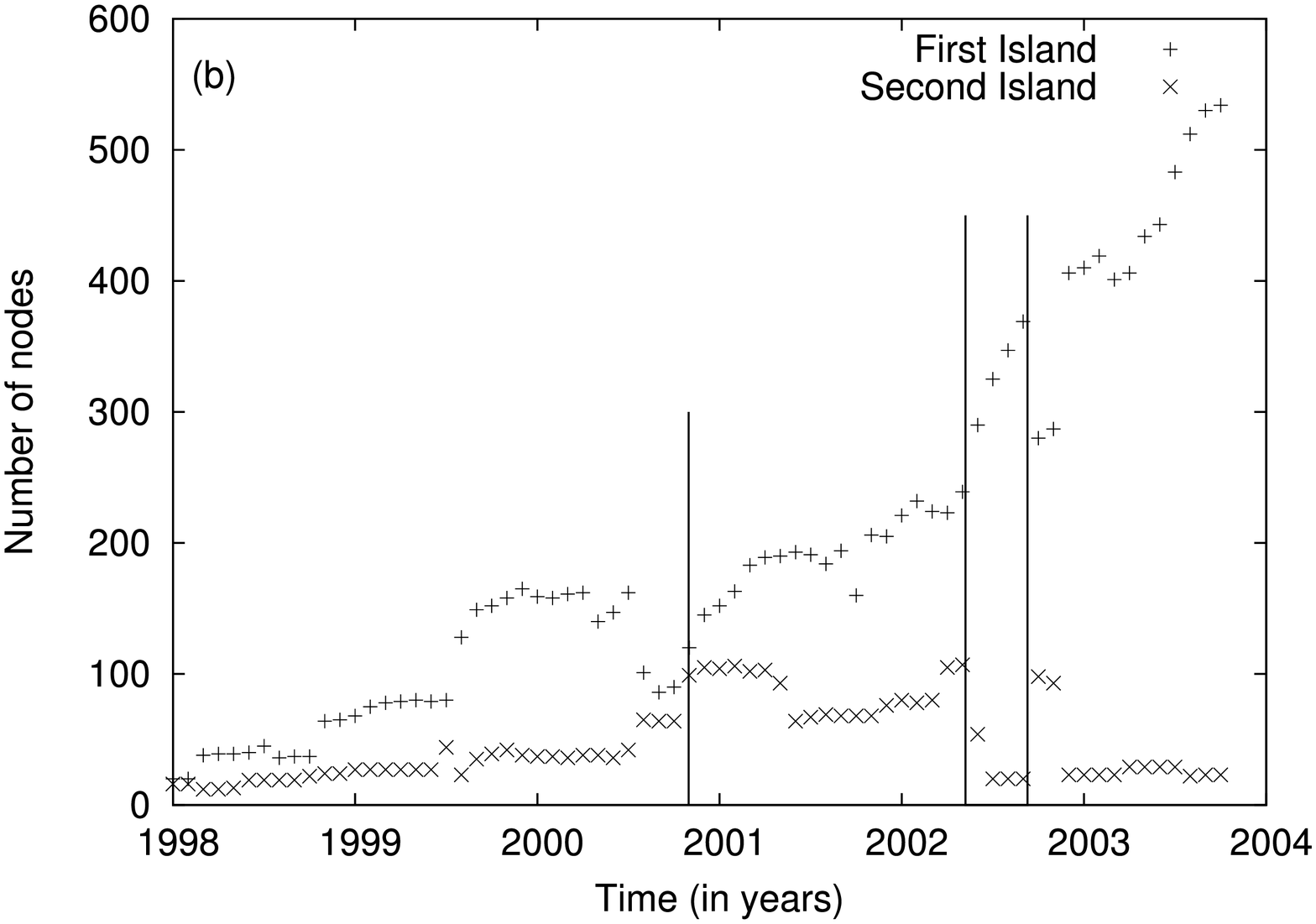}
\caption{\label{aaa} In (a), time evolution of the proportion of "network" links in the main island. In (b), time evolution of the number of nodes (scientists) in the two largest connected islands. The vertical lines point to the critical events occuring in the system (see main text).  }
\end{figure}

\begin{figure}
\includegraphics[width=2.2in]{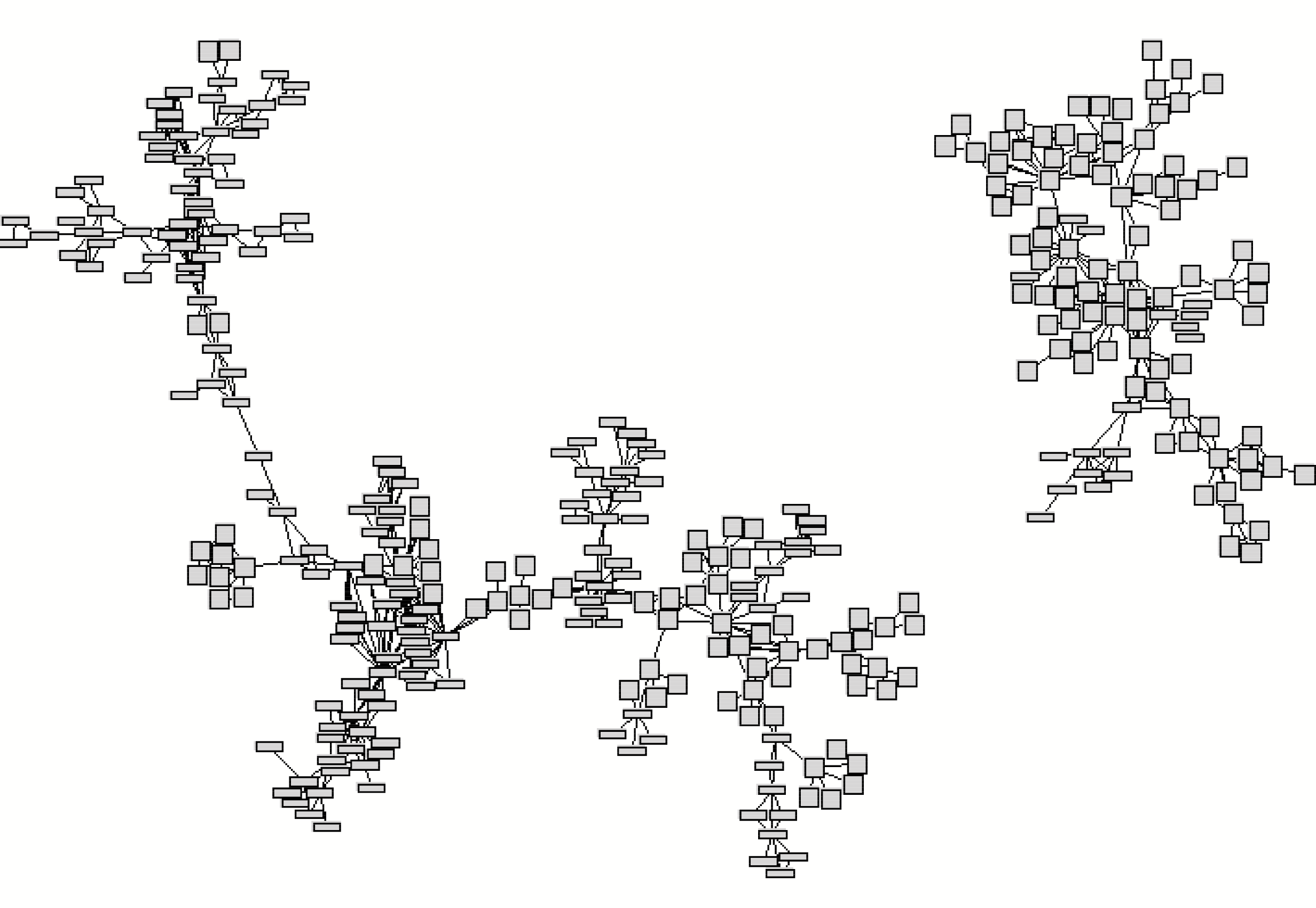}
\hspace{1cm}
\includegraphics[width=2.2in]{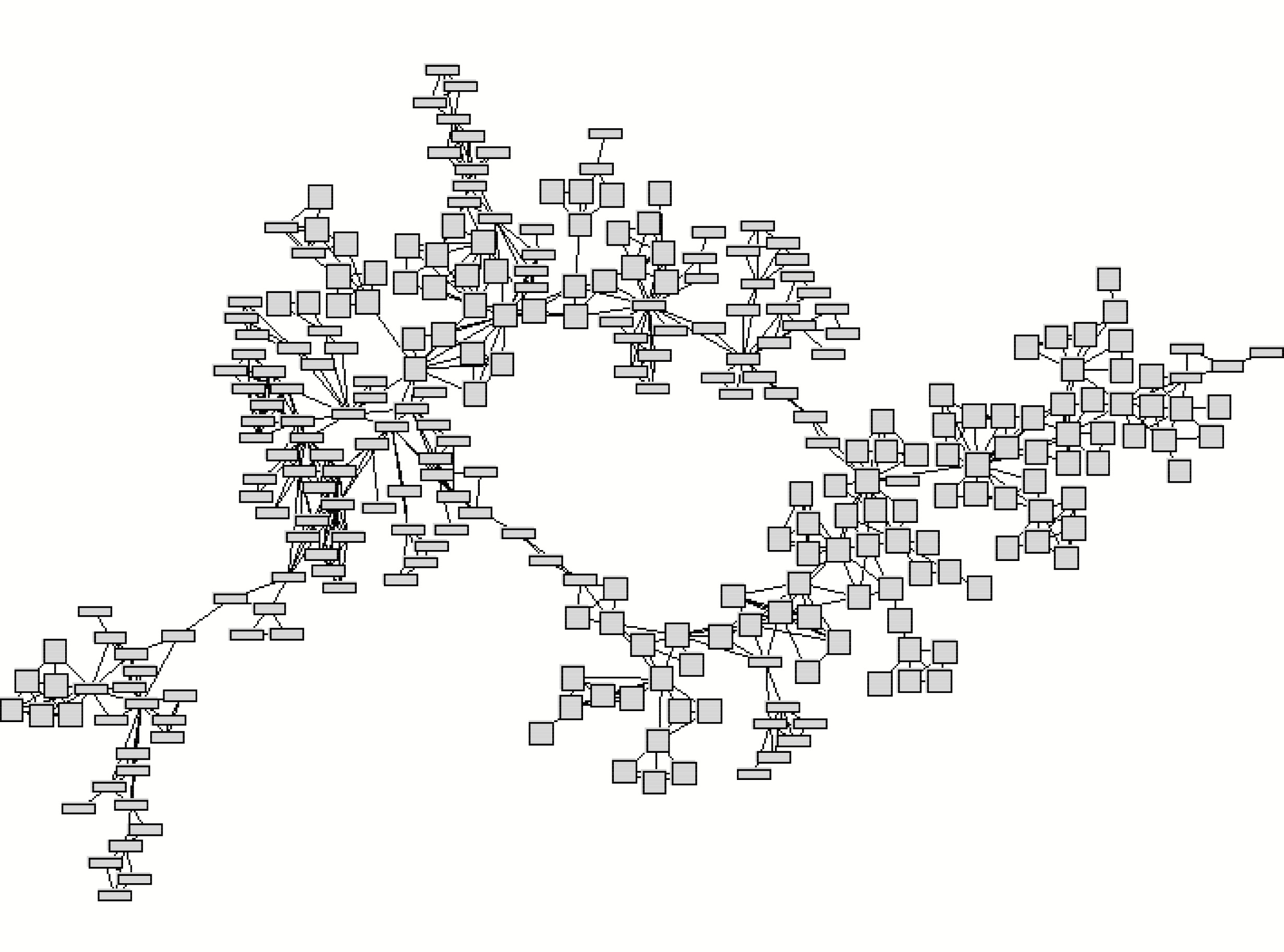}
\caption{\label{11} Merging of the two main sub-islands in the system. The left and right figure correspond respectively to September 2003 and January 2004. }
\end{figure}

In the remainder of this section, we focus on the time evolution of the above coupled networks. To do so, we consider overlapping time windows of 3 years, starting at July 1996, that we move forward in time by small intervals of 1 month. This method ensures a smooth time evolution of the different variables. Moreover, since now on, we characterize time windows by the date of the center value, e.g. we denote the interval [01/2002; 12/2004] by 07/2003.  In the following, we study in detail the properties of the main percolated island. The time evolution of the proportion of "network" links in this island exhibits  remarkable features (Fig.4a). There are obviously two important dates in the evolution, one around November 2000 where the system shows a first order transition from a "granular" state to a "network" state, and one around May 2002 where strong perturbations develop in the system. In order to find the origin of these critical behaviours, we have focused on the two largest connected islands in the system (Fig.4b).
Detailed analysis shows that the second largest island, centered around scientists like A.L. Barabasi, grows faster than the largest island, that is more focused on granular media and encompasses scientists like H. J. Herrmann.
Consequently, around November 2000, there is switch between the first and the second island, associated to discontinuities in the quantities describing the main island. This is therefore a first order transition, in analogy with equilibrium statistical mechanics. Around May 2002, another critical phenomena takes place, namely the two main islands merge together, thereby increasing the proportion of "granular" links in the largest island. It is interesting to note that, in August 2002, the two islands separate and recollide  two months later.
This bouncing and merging of the islands (Fig.5) is responsible for the fluctuations observed in (Fig.4a).

\begin{figure}
\hspace{-0.5cm}
\includegraphics[width=2.9in]{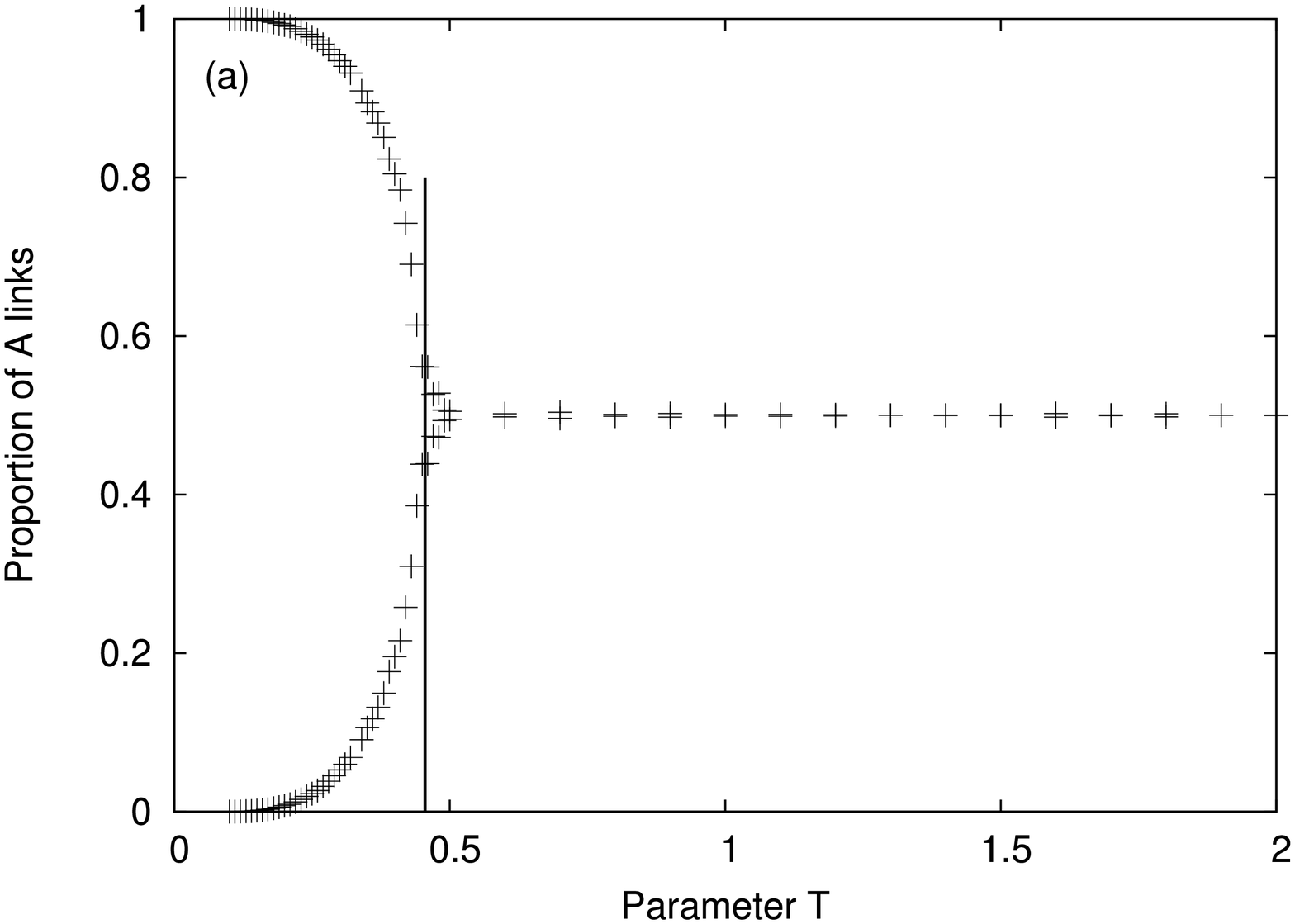}
\includegraphics[width=2.9in]{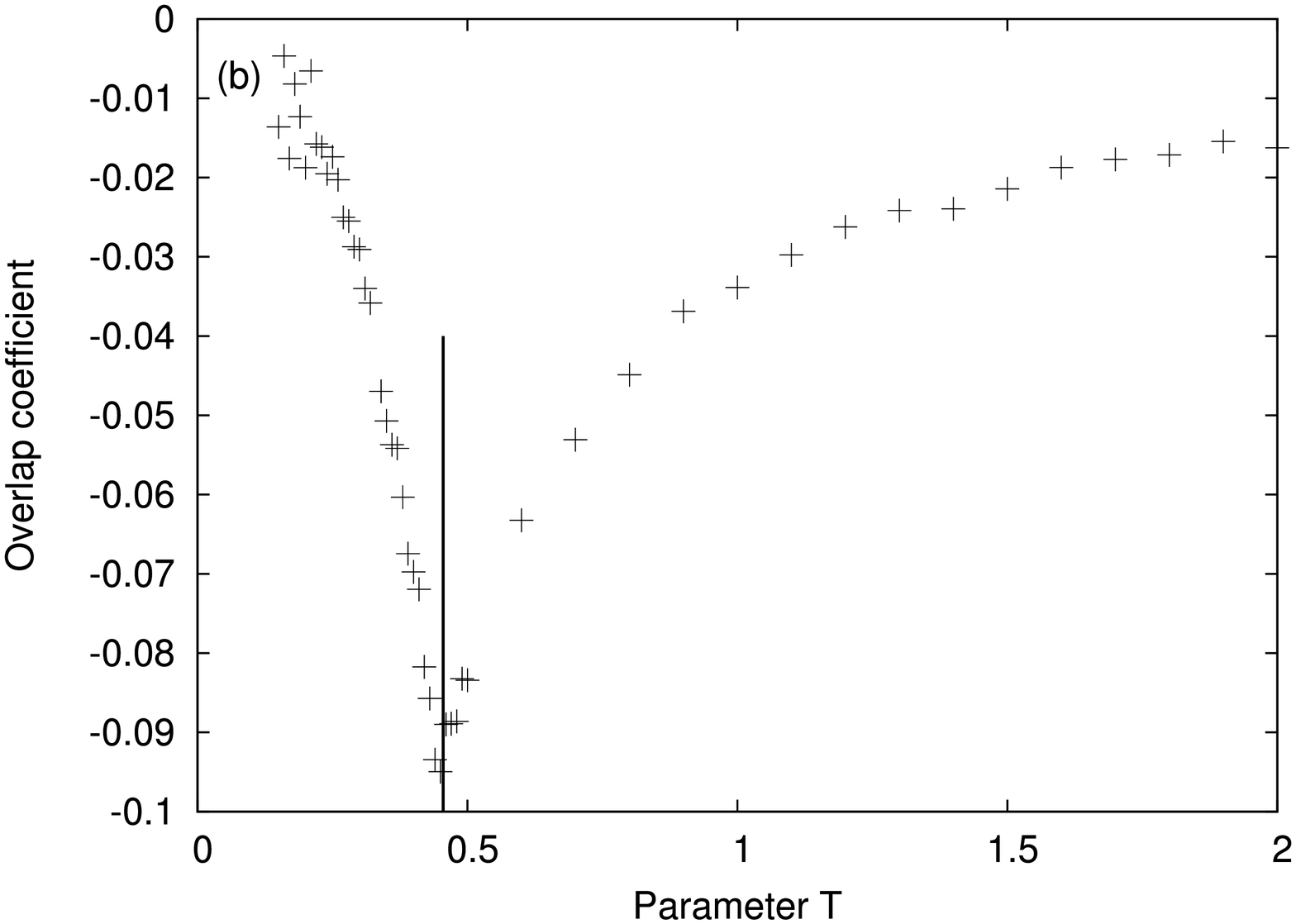}
\caption{\label{11} (a) Bifurcation diagram of CDIM, with 1000 agents, 10 links/agent, and $p_D=\frac{1}{2}$. (b) Bifurcation diagram for the overlap coefficient of the same system. In both figures, the vertical line points to the bifurcation point.}
\end{figure}

\section{Collaboration-Driven Ising Model}

In this last section, we introduce a very simple agent model, the Collaboration-Driven Ising Model (CDIM), that is based upon the above observations and is able to reproduce the self-organized emergence of phases in the system. To do so, we consider a stationary network, i.e. with constant number of scientists and of collaborations. There are 2 possible kinds of collaboration,  $A$ and $B$ ({\em"granular"}/{\em "network"}), between two scientists (no article written by 3, 4... scientists). The main assumptions follow. On the one hand, we assume that the state of the nodes is characterized by their previous collaborations, e.g. a scientist with a
 majority of $A$ collaborations is a $A$ scientist. Consequently, we neglect the influence of other internal variables (no spin) on the scientist future collaborations. On the other hand, we assume that scientists have a preference to work in their own field.
Practically, we consider a random network composed by $K$ nodes, and $N$ links. Initially, the $N$ links are randomly distributed as $A$ and $B$ links. At each time step, one link is removed and  a new link is added between 2 randomly selected nodes, $i$ and $j$.
The kind of the added link, $A$ or $B$, depends on the previous links of $i$ and of $j$. To model this mechanism, we calculate the proportions of links $A$ for $i$ and for $j$, $p_A^i=\frac{N^i_A}{N^i}$ and $p_A^j=\frac{N^j_A}{N^j}$, where $N_A^i$ and  $N^i$ denote respectively the number of links $A$ and the total number of links of the node $i$. These quantities measure the ability of $i$/$j$ to work in the field A. We define the pair ability to be the average $p_A^{ij}=\frac{(p_A^i + p_A^j)}{2}$, with $0 \leq p_A^{ij} \leq 1$. Therefore, if $p_A^{ij} > \frac{1}{2}$/$p_A^{ij} < \frac{1}{2}$, the selected pair should collaborate in the field $A$/$B$. We implement this mechanism with the probabilities $P_A=\frac{e^{\frac{p_A^{ij} -p_D}{T}}}{Z}$ and $P_B=\frac{e^{\frac{p_D - p_A^{ij} }{T}}}{Z}$ for the selected pair to collaborate in $A$/$B$. In these expressions, $Z$ is a normalizing constant, $T$ plays the role of a temperature and characterizes the curiosity of scientists, i.e. their ability to work in new fields. $p_D$ is a drift term, that breaks the internal symmetry, and mimics the external effect of political decisions on the dynamics. 

Simulations show that for high $T$, $A$ and $B$ links are randomly distributed in the network. Decreasing this parameter, structures develop in the system and lead to the emergence of separated phases for each scientific discipline, as those observed in figure \ref{mechanism}. Then, at some critical temperature $T_C$, a symmetry breaking takes place, associated with the spontaneous supremacy of one of the scientific disciplines $A$ or $B$. The bifurcation diagram (1000 agents, 10 links/agent) for the model without external field ($p_D=\frac{1}{2}$) is plotted in figure 6a, and confirms the analogy with a ferromagnetic transition. In order to characterize the interface between the 2 coupled networks, we  calculate the {\em overlap coefficient}, defined by $\Omega= \frac{<N^i_A N^i_B>}{<N^i_A><N^i_B>}-1$, where the averages are performed over the nodes $i$. By construction, $\Omega=0$ if the links A and B are independently distributed. In contrast, $\Omega<0$ indicates that few actors work in $A$ and $B$ simultaneously, i.e. the network is composed of well separated phases, where some scientists
ensure exchanges between the scientific communities. The bifurcation diagram for the overlap coefficient (Fig.6b) clearly shows that a decrease of $T$ is associated to a structuring of the network in separated phases. At the critical point $T_C$, a qualitative change takes place. Finally, let us stress that CDIM is analytically tractable in the mean field approximation. Indeed, by assuming that detailed balance takes place for the stationary solution, and that the fluctuations of the number of links/node are negligible, study of the dynamics master equation \cite{theo} show that $T^T_C=\frac{1}{2}$, independently of the number of nodes and links of the network \cite{prepa} . We have verified by simulations that $T_C$ remains in the vicinity of $\frac{1}{2}$, for a large number of parameters. Moreover, deviations from the theoretical value $|T_C-T^T_C|$ decrease when fluctuations of the number of links/node decrease, as expected.

\section{Conclusion}
Inter-connections between distinct scientific disciplines play a central role in primordial phenomena, including
the emergence of crises and trends in complex social networks, the diffusion of different topics in science and scientific avalanches, i.e. emergence of new research topics that rapidly attract large parts of the scientific community.
In this paper, we  focus empirically on this issue by studying data collected from the arXiv database, thereby highlighting the main mechanisms leading  to multi-disciplinarity, as well as a rich and complex phenomenology.
We also use the observations in order to
build a simple stationary model for connected multidisciplinary scientists. 
Qualitatively, its features are those of an Ising model for magnetic systems, even though its dynamics is driven by the collaboration links, and not by spin attached to nodes. For instance, the effect of the parameter $p_D$ is very similar to that of an external magnetic field, and leads to hysteresis and metastability \cite{prepa}.
It is worthwhile to stress that this preliminary model (CDIM) suffers limitations that avoid a quantitative comparison with the observed data. Indeed, CDIM does not incorporate mechanisms leading to power law degree distributions, and neglects many-author collaborations \cite{lambi2}, 
social effects (habits of authors to publish in close communities), non-stationary features,...
A generalization of CDIM that accounts for these effects is under progress.

{\bf Acknowledgements}
R.L. would like to thank A. Scharnhorst, I. Hellsten, K. Suchecki and J. Holyst for fruitful discussions.
This work 
has been supported by European Commission Project 
CREEN FP6-2003-NEST-Path-012864.


\begin{thebibliography}{0}
\bibitem{albert}
R. Albert and A-L Barabasi,{\em Rev. of Mod. Physics}, {\bf 74} (2002) 47

\bibitem{barabasi} 
A.-L. Barabasi and R. Albert, {\em Science} {\bf 286} (1999) 509 

\bibitem{internet}
R. Pastor-Satorras and A. Vespignani, {\em Evolution and Structure of the Internet : A Statistical Physics Approach}, Cambridge University Press, 2004

\bibitem{business}
S.-M. Yoon and K. Kim, arXiv physics/0503017

\bibitem{ecological}
R.J. Williams and N.D. Martinez {\em Nature} {\bf 404} (2000) 180

\bibitem{airplane}
A. Barrat, M. Barthelemy, R. Pastor-Satorras, and 
A. Vespignani, {\em Proc. Natl. Acad. Sci. USA} {\bf 101} (2004) 3747 

\bibitem{newman}  
M. E. J.  Newman, {\em Proc. Natl. Acad. Sci. USA},{\bf 98}, (2001), 404

\bibitem{newman3}
M. E. J. Newman, D. J. Watts, and S. H. Strogatz, {\em PNAS}, {\bf 99}, (2002), 2566

\bibitem{bara}
A.L. Barabasi,  H. Jeong, Z. Neda, E. Ravasz, A. Schubert and T. Vicsek, {\em Physica A}, {\bf 311},  (2002), 590

\bibitem{ramasco}  
J. J. Ramasco, S. N. Dorogovtsev and R. Pastor-Satorras, {\em Physical Review E}, {\bf 70},  (2004), 036106 

\bibitem{andrea}
E. Bruckner, W. Ebeling and A. Scharnhorst 
{\em Scientometrics} {\bf 18} (1990) 21


\bibitem{holyst}
J.A. Holyst, K. Kacperski and F. Schweitzer, 
{\em Annual Review of Comput. Phys.} {\bf 9} (2001) 253

\bibitem{levine}
A.J. Levine, T.C. Lubensky, {\em
Phys Rev E} {\bf 63} (2001) 041510

\bibitem{newman2b} 
M. E. J. Newman, {\em Phys. Rev. E} {\bf 64} (2001) 016132

\bibitem{site}
www.creen.org/rlambiot/sandNetwroks.html

\bibitem{newman2} 
M. E. J. Newman, S. H. Strogatz and D. J. Watts, {\em Phys. Rev. E}, {\bf 64}, (2001), 026118

\bibitem{scc}  
E.A. Variano, J.H. McKoy, H. Lipson, {\em Phys. Rev. Lett.}  {\bf 92}  {2004}  188701 

\bibitem{girvan1}
M. Girvan, and M. E. J. Newman {\em PNAS}, {\bf 99} (2002) 7821

\bibitem{theo}
R. B. Griffiths, C.-Y. Weng, and J. S. Langer, Phys. Rev. 149, 301 (1966)

\bibitem{prepa}
R. Lambiotte and M. Ausloos, in preparation


\bibitem{holyst2}
A. Aleksiejuk, J.A. Holyst and D. Stauffer, {\em  Physica A} {\bf 310} (2002) 260 

\bibitem{lambi2}
R. Lambiotte and M. Ausloos, arXiv physics/0507154







\end{thebibliography}
\end{document}